\documentclass[12pt]{iopart}

\usepackage{amssymb} 
\usepackage{soul}
\usepackage{xcolor}

\expandafter\let\csname equation*\endcsname\relax

\expandafter\let\csname endequation*\endcsname\relax

\usepackage{mathtools}

\usepackage{subfigure}
\usepackage{graphicx}
\usepackage[justification=centering]{caption}

\begin{document}

\title{Hidden symmetry in the biased Dicke Model}

\author{Xilin Lu$^1$, Zi-min Li$^1$, Vladimir V. Mangazeev$^1$ and Murray T. Batchelor$^{2,1,3}$}

\address{$^1$ Department of Theoretical Physics, Research School of Physics,
Australian National University, Canberra ACT 2601, Australia}
\address{$^2$ Mathematical Sciences Institute, Australian National University, Canberra ACT 2601, Australia}
\address{$^3$ Centre for Modern Physics, Chongqing University, Chongqing 40444, China}
\ead{murray.batchelor@anu.edu.au}
\vspace{10pt}

\begin{abstract}
The symmetry operators generating the hidden $\mathbb{Z}_2$ symmetry of the asymmetric quantum Rabi model (AQRM) at bias $\epsilon \in \frac{1}{2}\mathbb{Z}$ have recently been constructed by V. V. Mangazeev {\it et al.} [{\it J. Phys. A: Math. Theor.} {\bf 54}  12LT01 (2021)]. We start with this result to determine symmetry operators for the $N$-qubit generalisation of the AQRM, also known as the biased Dicke model, at special biases. We also prove for general $N$ that the symmetry operators, which commute with the Hamiltonian of the biased Dicke model, generate a $\mathbb{Z}_2$ symmetry.
\end{abstract}

\section{Introduction}

The Dicke model, originally proposed to investigate the super-radiant states of a quantum gas \cite{Dicke1954,HL1973}, 
describes the interaction between a single quantum mode and $N$ two-level atoms. 
The Dicke model has been widely studied (see, e.g., Refs.~\cite{Knight2004,Kirton2019,Reiter2020} and references therein).
Recently the dynamics equations of motion have been simplified by exploiting the $\mathbb{Z}_2$ symmetry of the model \cite{Reiter2020}.
The experimental realisation of the Dicke model opened the field of cavity quantum electrodynamics (QED) \cite{Knight2004,Kirton2019}. 
Due to the weak light-matter interaction within the cavity QED realisation, 
applications are, however, limited to small coupling strengths. 
This situation changed during the past two decades with the realisation of circuit QED \cite{Blais2004,Chiorescu2004}, 
where the atoms and light field are replaced by superconducting flux qubits and electronic circuits. 
As a result of using superconducting circuits, circuit QED is capable of realising the Dicke model with much larger light-matter coupling strengths 
$g/\omega$ close to or even larger than 1 \cite{Niemczyk2010,Yoshihara2018,Blais2020}. 
In addition, there is a natural generalisation of the Dicke model to the biased Dicke model by introducing biased superconducting flux qubits \cite{Ashhab2019}. 
These biases are experimentally easy to vary and are deeply related to the system properties.

One way to look at the biased Dicke model is to regard it as the $N$-qubit generalisation of the asymmetric quantum Rabi model (AQRM) \cite{Braak2011,Chen2012,XZBL}.
The AQRM is a member of the celebrated quantum Rabi model (QRM) \cite{Rabi1936, Braak2016, XZBL, Braak2019} family. 
It should be noted that the AQRM is different to the anisotropic QRM, 
for which the relative amplitude between the rotating/anti-rotating terms can be adjusted as discussed in Ref.~{\cite{Shen2017}}.
Similar to the QRM, the AQRM describes a single-mode bosonic field (cavity) interacting with one two-level atom (qubit) 
but in the presence of an extra bias term $\epsilon$. 
By taking the bias to zero the usual QRM is recovered. 
The two-qubit QRM \cite{Chilingaryan2013} has also been solved analytically \cite{Peng2014}, 
with different light-matter coupling strengths $g_1$  and $g_2$.

The QRM possesses a $\mathbb{Z}_2$ parity symmetry which is directly associated with the level crossings appearing in the energy spectrum. 
This parity symmetry is known to be broken at non-zero $\epsilon$. 
As a result, level crossings are not expected to be present in the spectrum of the AQRM. 
However, it was observed \cite{Braak2011,Li2015, W2017} that level crossings are restored when $\epsilon$ takes half-integer values.
Hence, at these values, there should be another symmetry present which has been called hidden symmetry. 
The origin of this hidden symmetry remained a puzzle until it was recently uncovered \cite{Mangazeev2021}, 
with a way to construct the corresponding symmetry operators, and the explicit expressions for the lowest few cases determined.

Hidden symmetry is also observed to be present in other AQRM-related models 
with broken $\mathbb{Z}_2$ symmetry \cite{Li2020, Xie2021}.
Examples include the biased versions of the two-qubit QRM, the two-mode QRM \cite{Zhang2013,Chilingaryan2015}, the
anisotropic QRM \cite{Tomka2014, Xie2014}, the Rabi-Stark model \cite{Grimsmo2013,Eckle2017,Xie2019} and the two-photon QRM \cite{Zhang2013}.
Similar to the AQRM, this symmetry is governed by the bias $\epsilon$ and an $\epsilon$-condition, 
which gives the values of $\epsilon$ for the crossings to occur \cite{Li2020,Xie2021}. 
As the $N$-qubit generalisation of the two-qubit AQRM, the biased Dicke model is also expected to possess similar hidden symmetry.

In this work, we identify the hidden symmetry in the biased Dicke model by determining the explicit expression for its symmetry operators 
and establishing its $\mathbb{Z}_2$ nature. 
The structure of this paper is organised as follows. 
In Section 2 we briefly review the hidden symmetry in the AQRM and the key ingredients for later use. 
In Section 3 we provide the $\epsilon$-condition for the biased $N$-qubit Dicke model based on the spectra 
and determine the symmetry operators $J$ for small $N$ using an ansatz from the AQRM. 
In Section 4 we conjecture the expression for the $J$-operator for arbitrary number of qubits and 
prove that it commutes with the biased Dicke model Hamiltonian. 
In Section 5 we prove the symmetry generated by $J$ is $\mathbb{Z}_2$ by showing that $J^2$ is a polynomial of the Hamiltonian. 
Detailed working is left to the appendices. 
In Section 6 we give concluding remarks.

\section{Hidden $\mathbb{Z}_2$ symmetry in the AQRM}

The AQRM generalises the QRM by adding an extra bias term, with Hamiltonian
\begin{equation}\label{AQRMHamiltonian}
   H_{\mathrm{AQRM}}= a^\dagger a+g \,\sigma_x(a^\dagger+a)+\Delta \, \sigma_z+\epsilon \,\sigma_x.
\end{equation}
Here $g$ is the coupling strength, $2\Delta$ is the level splitting and $\epsilon$ is the bias field. 
The frequency of the bosonic field $\omega$ is set to 1 for simplicity. 
Dropping the bias term in (\ref{AQRMHamiltonian}) reduces to $H_{\mathrm{QRM}}$.

One can check that when $\epsilon=0$, the parity operator
\begin{equation}
P=\sigma_z e^{\mathrm{i} \pi a^\dagger a}
\end{equation}
generates a $\mathbb{Z}_2$ symmetry
\begin{equation}
   [P, H_{\mathrm{QRM}}]=0, \qquad P^2=\mathbb{I}, 
\end{equation}
with $\mathbb{I}$ the identity operator.
This symmetry separates the eigenstates of the QRM Hamiltonian into two sectors, 
labelled by the corresponding positive or negative eigenvalues of $P$. 
The crossings in the spectrum are then between pairs from different sectors \cite{Braak2019}.

As mentioned earlier, these crossings split as the bias $\epsilon$ is varied, but are recovered when $\epsilon$ is a half-integer. 
The symmetry operator $J$ corresponding to this observation has recently been found \cite{Mangazeev2021}, 
in the form of $2\times 2$ matrices with entries polynomials in $a$ and $a^\dagger$. 
The degree of these polynomials is equal to $M$, where $\epsilon={M}/{2}$, and their coefficients can be determined by solving a system of partial differential equations. 
This kind of expression was later proven \cite{Reyes-Bustos2021} to be unique (up to scaling). 
The case $M=0$ simply gives the parity operator $P$ while the case $M=1$ reads 
\begin{equation}
    J_{\epsilon=1/2}=\mathcal{P}\begin{pmatrix}
    a^\dagger-a+2g+\frac{\Delta}{g} & a^\dagger+a\\
    -a^\dagger-a & a-a^\dagger-2g+\frac{\Delta}{g}
    \end{pmatrix}.
\end{equation}
Here we define 
\begin{equation}
    \mathcal{P}=e^{ \mathrm{i} \pi a^\dagger a},
\end{equation}
which anti-commutes with bosonic operators $a^\dagger$ and $a$:
\begin{equation}
   a^\dagger \mathcal{P}=-\mathcal{P}a^\dagger, \qquad a \mathcal{P}=-\mathcal{P}a.
\end{equation}
The expressions for the $J$ operators quickly become tedious with increasing $M$, 
hence we only give this first non-trivial case for later use.

The $J$ operators indeed generate the $\mathbb{Z}_2$ symmetry because they square to polynomials of the AQRM Hamiltonian, 
where the degree of the polynomials are also equal to $M$. 
Therefore, similar to the parity operator, they divide the eigenstates of $H_{AQRM}$ into two sets and crossings 
can occur between states from different sets. The $M=1$ example gives 
\begin{equation}
    J_{\epsilon=1/2}^2=4H_{\mathrm{AQRM}}+4g^2+\frac{\Delta^2}{g^2}+2.
\end{equation}

Knowing the form of $J_{\mathrm{AQRM}}$, our general strategy is to investigate $J$ operators for other related models by making an 
ansatz similar to $J_{\mathrm{AQRM}}$.

\section{Biased Dicke model with small $N$}

The biased $N$-qubit Dicke model has Hamiltonian
\begin{equation}\label{BDMHamiltonian}
   H^{Nq}=a^\dagger a+ \sum_{i=1}^N (\Delta_i \, \sigma^z_i+\epsilon_i \, \sigma_i^x)+(a^\dagger+a)\sum_{i=1}^N g_i \, \sigma_i^x,
\end{equation}
with $N$ sets of parameters $\{\Delta,g,\epsilon\}$. 
Here we adapt the notation from the quantum inverse scattering method (QISM) \cite{Korepin1993} 
for which the subscript of an operator indicates the location in the tensor vector space that it acts on. 
For example, with $\mathbb{I}$ again the identity operator, 
$\sigma_i^x$ is the Pauli matrix $\sigma_x$ acting on the $i$-th qubit: 
\begin{equation}
    \sigma_i^x :=  \mathbb{I} \otimes \cdots \otimes \underbrace{\sigma_x}_{i\text{-th}} \otimes \cdots \otimes \mathbb{I}.
\end{equation}

We investigate the hidden symmetry for a small number of qubits in this section, namely $N=2,3,4$. 
According to the behaviour of their energy spectra, the $\epsilon$-condition here is $\Delta_i=\Delta,g_i=g$ for all $i$, 
with only one $\epsilon_k=\frac{1}{2}$ and $\epsilon_i=0$ for $i\neq k$. 
Surprisingly, crossings are not observed for higher $\epsilon_k$, such as 1 or $\frac{3}{2}$. 
We consider this fact as a result of the extra term in the $J$ operator associated with the $B$-matrix which will be introduced soon. 

With the identical qubits condition, $\Delta_i=\Delta, g_i=g,\epsilon_i=\epsilon$ for all $i$, 
the permutation symmetry $S_N$ is present in the biased Dicke model. 
This means that for the $\epsilon$-condition stated above, 
we can always observe crossings in the spectra when $N>2$ due to the remaining {$S_{N-1}$} symmetry, 
while the additional $Z_2$ symmetry results in extra crossings in the spectra. 
Therefore, we will illustrate the case $N=2$ below to exclude crossings due to permutation symmetry.

\begin{figure}[h]
    \centering
    \subfigure[$g_2=g_1$, $\epsilon_1=0$]{\label{fig:spectra1}
    \includegraphics[width=0.4\textwidth]{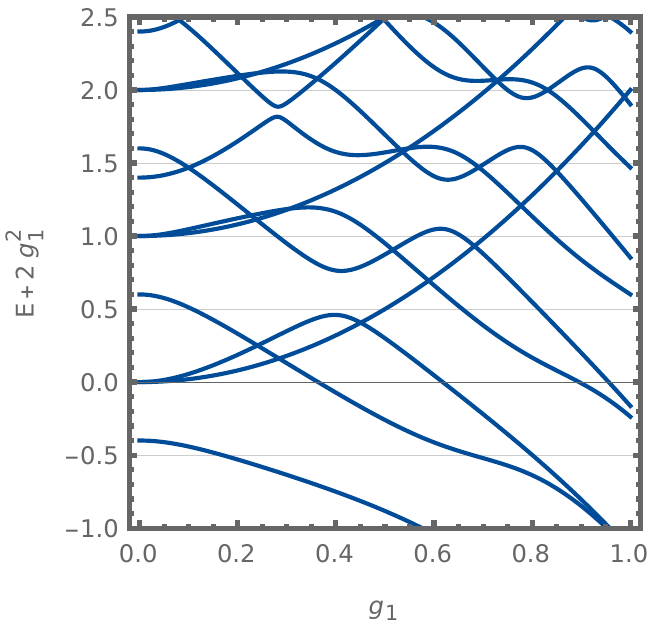}}
    \subfigure[$g_2=g_1$, $\epsilon_1=\frac{1}{2}$]{\label{fig:spectra2}\includegraphics[width=0.4\textwidth]{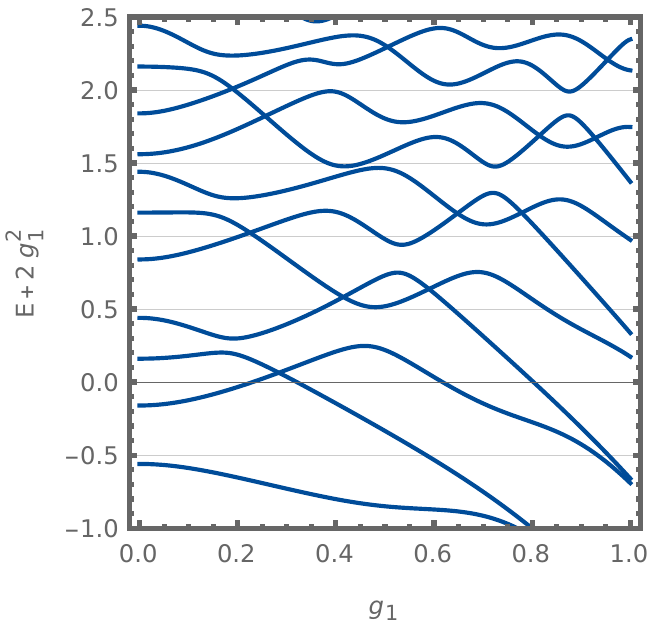}} 
    \subfigure[$g_2=g_1$, $\epsilon_1=1$]{\label{fig:spectra3}\includegraphics[width=0.4\textwidth]{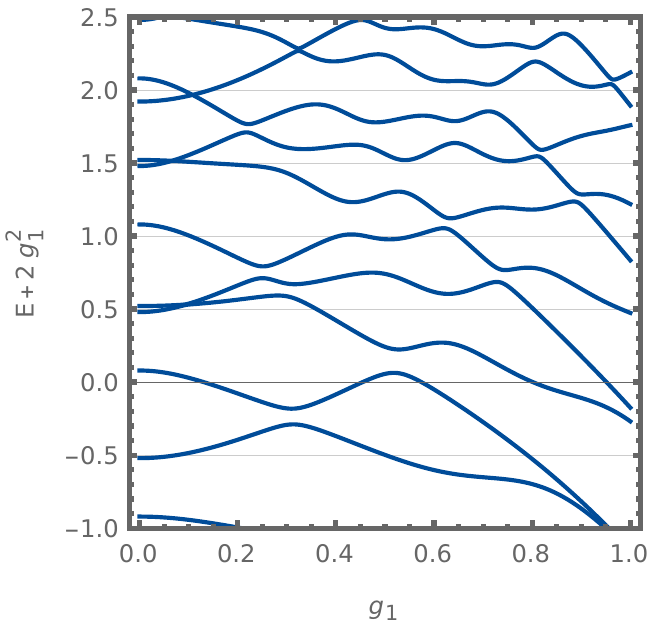}}
    \subfigure[$g_2=1.2 g_1$, $\epsilon_1=\frac{1}{2}$]{\label{fig:spectra4}\includegraphics[width=0.4\textwidth]{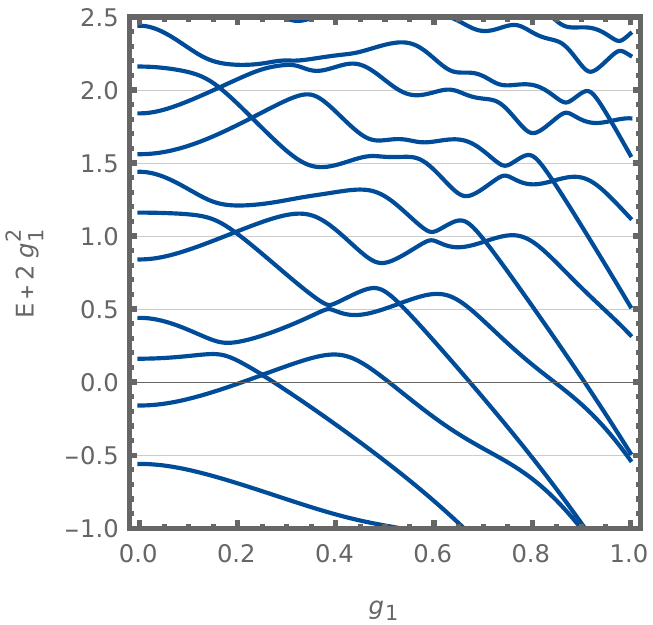}}
   \caption{Spectrum of the two-qubit biased Dicke model as a function of the coupling $g_1$ with $\Delta_1=\Delta_2=0.7$, $\epsilon_2=0$ and (a) $g_2=g_1$, $\epsilon_1=0$ (b) $g_2=g_1$, $\epsilon_1=\frac{1}{2}$ (c) $g_2=g_1$, $\epsilon_1=1$ (d) $g_2=1.2 g_1$, $\epsilon_1=\frac{1}{2}$. }
    \label{fig:spectra}
\end{figure}

Some example spectra for the $N=2$ case are shown in Figure \ref{fig:spectra}. 
Firstly, we can observe that crossings appear in Figure \ref{fig:spectra1}, where $\Delta_1=\Delta_2$, $\epsilon_1=\epsilon_2=0$ and $g_1=g_2$. 
The model is not biased in this case hence the parity symmetry is present, similar to the QRM. 
Next, we start to apply the bias $\epsilon_1$. 
There is no crossing until $\epsilon_1=\frac{1}{2}$, which corresponds to the $M=1$ case of the hidden symmetry in the AQRM. 
As shown in Figure \ref{fig:spectra2}, crossings appear hence the hidden symmetry is present. 
Then, in Figure \ref{fig:spectra3} we further deform the bias to $\epsilon_1=1$, which corresponds to the $M=2$ case. 
Surprisingly, there is no crossing present (the apparent crossings are seen to be avoided crossings when zooming-in), 
and similar observations were found for larger $M$ cases. 
Finally, in Figure \ref{fig:spectra4} we change the ratio between $g_1$ and $g_2$ while keeping $\epsilon_1=\frac{1}{2}$. 
There is again no real crossing under closer inspection, hence we need $g_1=g_2$ in the $\epsilon$-condition, 
and from the same observation we need $\Delta_1=\Delta_2$.

To make calculations easier, we can set $k=1$ without loss of generality. Hence the Hamiltonian (\ref{BDMHamiltonian}) becomes
\begin{equation}
   \tilde{H}^{Nq}=a^\dagger a+\Delta \sum_{i=1}^N \sigma^z_i+g(a^\dagger+a)\sum_{i=1}^N \sigma_i^x+\frac{1}{2}\sigma_1^x.
\end{equation}
Then, similar to the AQRM case, we assume the $J$ operators take the form
\begin{equation}\label{ansatz}
    J^{Nq}_M= \mathcal{P}\mathcal{Q}^{Nq}_M(a^\dagger,a),
 \end{equation}
where $\mathcal{Q}^{Nq}_M(a^\dagger,a)$ is a $2^N \times 2^N$ matrix acting in the the space $(\mathbb{C}^2)^{\otimes N}\otimes \mathcal{H}$ with entries
\begin{equation}
 (\mathcal{Q}_M(a^\dagger,a))_{\alpha,\beta}=\sum_{i,j=1}^M Q^{(\alpha,\beta)}_{i,j}(a^\dagger)^i a^j.
\end{equation} 
Here $\mathcal{H}$ is the Fock space and $M=2\epsilon$ is again the order of polynomials.

From the $\epsilon$-condition above we only have $\epsilon_k=\frac{1}{2}$, 
then $M=1$ and elements of $\mathcal{Q}$ are monomials of $a^\dagger$ and $a$. 
Therefore, we can drop the subscript 1 and calculate the $J^{Nq}$ operators for small $N$ values by assuming $J^{Nq}$ commutes with $\tilde{H}^{Nq}$. 
The two-qubit case $J^{2q}$ reads
\begin{equation}
    J^{2q}=J^{1q}_1 \sigma_2^z|_{\Delta,g\rightarrow 2\Delta,2g}+4g\mathcal{P}B_{12},
\end{equation}
where $J^{1q}=J_{\epsilon=1/2}$ is the $J$ operator for the normal 1-qubit AQRM and 
$B_{ij}$ is a constant matrix acting on the tensor product of spaces $i$ and $j$ defined as
\begin{equation} \label{BDef}
   B_{ij}\equiv \begin{pmatrix}
    0&0&0&0\\
    0&1&-1&0\\
    0&-1&1&0\\
    0&0&0&0
    \end{pmatrix}_{i,j}.
\end{equation}

Similarly, for the 3- and 4-qubit cases we have
\begin{align}
    &J^{3q}=J^{1q}_1 \sigma_2^z \sigma_3^z|_{\Delta,g\rightarrow 3\Delta,3g}+4g\mathcal{P}(B_{12}\sigma_3^z+B_{13}\sigma_2^z),\\
    &J^{4q}=J^{1q}_1 \sigma_2^z \sigma_3^z\sigma_4^z|_{\Delta,g\rightarrow 4\Delta,4g}+4g\mathcal{P}(B_{12}\sigma_3^z\sigma_4^z+B_{13}\sigma_2^z\sigma_4^z+B_{14}\sigma_2^z\sigma_3^z).
\end{align}
Now we begin to see some patterns in these expressions. 
We will make a conjecture based on these patterns and prove it in the next section.

\section{Symmetry operator $J^{Nq}$ for general $N$}

From the observation in the last section, we find the operator $J^{Nq}$ appears to be made up of two parts. 
The first part is the $J^{1q}$ operator with tensor products of the $\sigma_z$ matrices in all other spaces, 
and the second part is $\mathcal{P}$ times the sum of $B$ matrices, 
mixing the asymmetric qubit with a symmetric qubit, and then tensor products with  $\sigma_z$ on all other qubits. 
Based on these facts, we propose and prove the following theorem.

\textbf{Theorem 1}. {\it For the $N$-qubit biased Dicke model, with $\Delta_i=\Delta,g_i=g$ for all $i$, 
$\epsilon_k=\frac{1}{2}$ and $\epsilon_j=0$ for $j\neq k$, there exists an operator $J^{Nq}$ such that it commutes with the Hamiltonian $H^{Nq}$ (\ref{BDMHamiltonian}) and admits the form
\begin{equation}\label{JoperatorDef}
    J^{Nq}=J^{1q}_k|_{\Delta,g\rightarrow N\Delta,Ng}\prod_{j\neq k}^{N}\sigma_j^z+4g\mathcal{P}\sum_{j\neq k}^{N}B_{kj}\prod_{m\neq j,k}^{N}\sigma_m^z,
\end{equation}
where the  indices run from 1 to $N$.}

A proof of this theorem is given in Appendix A. 
The line of proof is to show that the commutator $\left[J^{(N+1)q},H^{(N+1)q} \right]$ vanishes if $\left[J^{Nq},H^{Nq} \right] =0$. 
Since we already know \cite{Mangazeev2021} that $\left[J^{1q},H^{1q} \right] =0$ the theorem follows by induction.
The operator $J^{Nq}$ defined in (\ref{JoperatorDef}) commutes with the Hamiltonian (\ref{BDMHamiltonian}) 
for arbitrary number of qubits and hence generates a symmetry of the $N$-qubit biased Dicke model. 
The next step is to show that the symmetry generated by $J^{Nq}$ is a $\mathbb{Z}_2$ symmetry.

\section{Hidden $\mathbb{Z}_2$ symmetry in the biased Dicke model}

We show in this section that the symmetry generated by the $J$ operator (\ref{JoperatorDef}) is a $\mathbb{Z}_2$ symmetry, 
analogous to the 1-qubit AQRM case. 
To see this, we can start by observing that the operator $J^{2q}$ satisfies the same quadratic relation with the Hamiltonian as $J^{1q}$, namely
\begin{equation}
    (J^{2q})^2=4H^{2q}+\frac{\Delta^2}{g^2}+16g^2+2.
\end{equation}
Recall that for $J^{1q}$ \cite{Mangazeev2021},
\begin{equation}
    (J^{1q})^2=4H^{1q}+\frac{\Delta^2}{g^2}+4g^2+2.
\end{equation}
It follows that by doubling $\Delta$ and $g$ we can arrive at the relation for $J^{2q}$. 

However, the situation is more complicated for larger $N$. 
We calculate, for example, the 3- and 4-qubit cases (with $\epsilon_1=\frac{1}{2}$ and other $\epsilon$'s zero) for further insights. 
For these cases we find 
\begin{align}
   &(J^{3q})^2=4H^{3q}+\frac{\Delta^2}{g^2}+36g^2+2-16g^2 B_{23}, \\
   &(J^{4q})^2=4H^{4q}+\frac{\Delta^2}{g^2}+64g^2+2-16g^2 (B_{23}+B_{24}+B_{34}). 
\end{align}
We see that the $B$-matrices again appear but are not associated with the first (asymmetric) qubit as before. 
From these patterns, we propose and prove our second theorem.

\textbf{Theorem 2}. {\it The $J^{Nq}$ operator defined by (\ref{JoperatorDef}) with Hamiltonian $H^{Nq}$ defined by (\ref{BDMHamiltonian}) 
satisfies the quadratic relation 
\begin{equation}\label{quadraticRelationJandH}
    (J^{Nq})^2=4H^{Nq}+\frac{\Delta^2}{g^2}+2+4N^2g^2-16g^2\sum_{j\neq k}^N\sum_{\substack{m>j\\
    m\neq k}}^N B_{jm}.
\end{equation}
}

A proof of this theorem is given in Appendix B, again by using induction. 
A step in the proof relies on noticing the properties 
\begin{equation} \label{prop1}
 \sigma_i^z\sigma_{j}^zB_{ij}=B_{ij}\sigma_i^z\sigma_{j}^z=-B_{ij}  
\end{equation}
and
\begin{equation} \label{prop2}
   (B_{ij})^2=2B_{ij}. 
\end{equation}
These properties can be verified by direct computation using definition ({\ref{BDef}}). 
We also use the $B$-matrix property
\begin{equation} \label{prop3}
    B_{ij}\sigma_j^z\sigma_k^z B_{ik}+B_{ik}\sigma_k^z\sigma_j^z B_{ij}=B_{ij}+B_{ik}-B_{jk}.  
\end{equation}
To prove this property, we rewrite the $B$-matrix in terms of Pauli matrices, 
\begin{equation} \label{Bmatrices}
    B_{ij}=\frac{1}{2}(\mathbb{I}-\sigma_i^x\sigma_j^x-\sigma_i^y\sigma_j^y-\sigma_i^z\sigma_j^z), 
\end{equation}
and expand both sides to see that they are equal. 

In addition to the commutation relation $\left[J^{Nq},H^{Nq} \right]=0$, 
the quadratic relation (\ref{quadraticRelationJandH}) in terms of $H^{Nq}$ is sufficient to establish  
the existence of $\mathbb{Z}_2$ symmetry within the biased Dicke model.

\section{Conclusion}

In this work, we found the $\epsilon$-condition for the hidden $\mathbb{Z}_2$ symmetry 
in the biased Dicke model $H^{Nq}$ (\ref{BDMHamiltonian}) and determined the explicit expression (\ref{JoperatorDef}) 
for the corresponding symmetry operator $J^{Nq}$. 
This was achieved by using the ansatz (\ref{ansatz}) based on our recent results for the AQRM \cite{Mangazeev2021}. 
We proved that the $J^{Nq}$ operator commutes with the Hamiltonian $H^{Nq}$ (\ref{BDMHamiltonian}) and has a quadratic relation (\ref{quadraticRelationJandH}) in terms of $H^{Nq}$,  
thereby establishing the existence of $\mathbb{Z}_2$ symmetry within the biased Dicke model.

We note that the matrices $B$ (\ref{Bmatrices}) appear in both the expression for $J^{Nq}$ (\ref{JoperatorDef}) and the quadratic relation (\ref{quadraticRelationJandH}). 
This appears to be associated with the fact that only the lowest order of hidden symmetry ($\epsilon_k=\frac{1}{2}$) is present in the biased Dicke model. 
It is also interesting to observe that the product $\mathcal{P}B$ 
is a special case of the permutation symmetry \cite{Peng2014} in the two-qubit QRM.
Our result for the hidden symmetry of the biased Dicke model in the $N=2$ case 
begs the question if the analytic solution of the two-qubit QRM  \cite{Peng2014} 
can be extended to include a bias term.

Finally, we note that the ansatz (\ref{ansatz}) works surprisingly well, it not only leads to the $J$ operators for the biased Dicke model, 
but is also capable of providing solutions for other AQRM-related models. 
For example, we have also used this ansatz to determine the $J$ operators for the anisotropic AQRM and the asymmetric Rabi-Stark model \cite{Lu2021}. 
The intertwining of the operator $\mathcal{P}$ in $J$ means that this hidden symmetry is a kind of generalisation of the parity symmetry. 
We believe that further understanding of this symmetry in the biased Dicke and other related models can be obtained by studying the physical meaning of the $J$ operators.

\ack

The authors thank Daniel Braak for helpful comments on underlying symmetries.
This work has been supported by the Australian Research Council Discovery Projects DP170104934 and DP180101040.

\appendix

\section{Proof of Theorem 1}

Without loss of generality, we can prove the statement for $k=1$ and then permute qubits to the spaces one desires. 
We prove the case $k=1$ using the following induction process. 

From Section 2 we know that 
\begin{equation}
\left[J^{1q},H^{1q}\right]=0.    
\end{equation}
Now, we want to prove the ($N$+1)-qubit case assuming the $N$-qubit case is true. 
Then we have 
\begin{equation}
\begin{split}
    \left[ J^{Nq},H^{Nq} \right]=&[J^{1q}_1|_{\Delta,g\rightarrow N\Delta,Ng}\prod_{j=2}^N \sigma_j^z+4g\mathcal{P}\sum_{j=2}^N B_{1j}\prod_{m\neq j}^{N}\sigma_m^z, \\
    &a^\dagger a+\Delta \sum_{i=1}^N \sigma^z_i+g(a^\dagger+a)\sum_{i=1}^N \sigma_i^x+\frac{1}{2}\sigma_1^x]=0
\end{split}
\end{equation}
for the $N$-qubit case. 
By direct evaluation one can verify that the part of $J^{Nq}$ with $B$-matrices commutes with the parts of the Hamiltonian except the bias term 
\begin{equation}
    \left[4g\mathcal{P}\sum_{j=2}^N B_{1j}\prod_{m\neq j}^{N}\sigma_m^z, \, a^\dagger a+\Delta \sum_{i=1}^N \sigma^z_i+g(a^\dagger+a)\sum_{i=1}^N \sigma_i^x \right]=0. 
\end{equation}
We thus arrive at the equation
\begin{equation}\label{dickeNqubit}
    \begin{split}
    & \left[ 4g\mathcal{P}\sum_{j=2}^N B_{1j}\prod_{m\neq j}^{N}\sigma_m^z,\, \frac{1}{2}\sigma_1^x \right] \\
    &=-\left[J^{1q}_1|_{\Delta,g\rightarrow N\Delta,Ng}\prod_{j=2}^N \sigma_j^z, \, a^\dagger a+\Delta \sum_{i=1}^N \sigma^z_i+g(a^\dagger+a)\sum_{i=1}^N \sigma_i^x+\frac{1}{2}\sigma_1^x \right],
    \end{split}
\end{equation}
which will be the key for the next step.

Now we look at the ($N$+1)-qubit case. 
Again, we calculate the commutator between $J^{(N+1)q}$ and the Hamiltonian
\begin{equation}\begin{split}
 \left[J^{(N+1)q},H^{(N+1)q} \right]=[&J^{1q}_1|_{\Delta,g\rightarrow (N+1)\Delta,(N+1)g}\prod_{j=2}^{N+1} \sigma_j^z+4g\mathcal{P}\sum_{j=2}^{N+1} B_{1j}\prod_{m\neq j}^{N+1}\sigma_m^z,\\
    &a^\dagger a+\Delta \sum_{i=1}^{N+1} \sigma^z_i+g(a^\dagger+a)\sum_{i=1}^{N+1} \sigma_i^x+\frac{1}{2}\sigma_1^x].
\end{split}
\end{equation}
Similar to the $N$-qubit case, we can break the $J$ operator into the sum of two terms, 
where one of them commutes with the Hamiltonian apart from $\sigma_1^x$. 
In this way, the above expression can be rewritten as
\begin{equation}\label{dickeNp1qubit}\begin{split}
 [J^{(N+1)q}&,H^{(N+1)q}]=\left[4g\mathcal{P}\sum_{j=2}^{N+1} B_{1j}\prod_{m\neq j}^{N+1}\sigma_m^z,\frac{1}{2}\sigma_1^x \right]\\
 &+\left[J^{1q}_1|_{\Delta,g\rightarrow (N+1)\Delta,(N+1)g}\prod_{j=2}^{N+1} \sigma_j^z, \, a^\dagger a+\Delta \sum_{i=1}^{N+1} \sigma^z_i+g(a^\dagger+a)\sum_{i=1}^{N+1} \sigma_i^x+\frac{1}{2}\sigma_1^x \right].
\end{split}
\end{equation}

Now we can analyse these two commutators separately. 
For the first one, we have  
\begin{equation}\label{dicketerm1}\begin{split}
 \left[4g\mathcal{P}\sum_{j=2}^{N+1} B_{1j}\prod_{m\neq j}^{N+1}\sigma_m^z,\frac{1}{2}\sigma_1^x\right]
 &=\left[4g\mathcal{P}\sum_{j=2}^{N} B_{1j}\prod_{m\neq j}^N\sigma_m^z \sigma_{N+1}^z+4g\mathcal{P}B_{1,N+1}\prod_{j=2}^N \sigma_j^z,\frac{1}{2}\sigma_1^x \right]\\
 &=\left[4g\mathcal{P}\sum_{j=2}^{N} B_{1j}\prod_{m\neq j}^N\sigma_m^z,\frac{1}{2}\sigma_1^x\right]\sigma_{N+1}^z+\left[4g\mathcal{P}B_{1,N+1}\prod_{j=2}^N \sigma_j^z,\frac{1}{2}\sigma_1^x \right].
\end{split}
\end{equation}
According to (\ref{dickeNqubit}), the first term in (\ref{dicketerm1}) is equal to 
\begin{equation}
    -\left[J^{1q}_1|_{\Delta,g\rightarrow N\Delta,Ng}\prod_{j=2}^N \sigma_j^z, \, a^\dagger a+\Delta \sum_{i=1}^N \sigma^z_i+g(a^\dagger+a)\sum_{i=1}^N \sigma_i^x+\frac{1}{2}\sigma_1^x \right]\sigma_{N+1}^z. 
\end{equation}
The second term can be determined by direct calculation, with
\begin{equation}
   \begin{split}
   [B_{1,N+1},\sigma_1^x]=&\left[\begin{pmatrix}
    0&0&0&0\\
    0&1&-1&0\\
    0&-1&1&0\\
    0&0&0&0
   \end{pmatrix}_{1,N+1},\begin{pmatrix}
    0&0&1&0\\
    0&0&0&1\\
    1&0&0&0\\
    0&1&0&0
   \end{pmatrix}_{1,N+1}\right]= \mathrm{i} (\sigma_1^z\sigma_{N+1}^y-\sigma_1^y\sigma_{N+1}^z).
   \end{split} 
\end{equation}
For the first commutator in (\ref{dickeNp1qubit}) it follows that we have the  simplified form 
\begin{equation}\label{dicketerm1finalform}\begin{split}
 \left[4g\mathcal{P}\sum_{j=2}^{N+1} B_{1j}\prod_{m\neq j}^{N+1}\sigma_m^z,\frac{1}{2}\sigma_1^x\right]
 &=-\left[J^{1q}_1|_{\Delta,g\rightarrow N\Delta,Ng}\prod_{j=2}^N \sigma_j^z,a^\dagger a+\Delta \sum_{i=1}^N \sigma^z_i+g(a^\dagger+a)\sum_{i=1}^N \sigma_i^x \right. \\
 & ~~~ \left. +\frac{1}{2}\sigma_1^x \right]\sigma_{N+1}^z
 +2g\mathcal{P}\prod_{m\neq j}^{N}\sigma_m^z \mathrm{i} (\sigma_1^z\sigma_{N+1}^y-\sigma_1^y\sigma_{N+1}^z).
\end{split}
\end{equation}

Now we can simplify the second commutator in (\ref{dickeNp1qubit}) by separating terms associated with the ($N+1$)-th qubit. 
Also by noticing that 
\begin{equation}
    J^{1q}_1|_{\Delta,g\rightarrow (N+1)\Delta,(N+1)g}=J^{1q}_1|_{\Delta,g\rightarrow N\Delta,Ng}+2g\mathcal{P}\sigma_1^z,
    \end{equation}
we have
\begin{equation}\label{dicketerm2}\begin{split}
& \left[J^{1q}_1 |_{\Delta,g\rightarrow (N+1)\Delta,(N+1)g}\prod_{j=2}^{N+1} \sigma_j^z, \, a^\dagger a+\Delta \sum_{i=1}^{N+1} \sigma^z_i+g(a^\dagger+a)\sum_{i=1}^{N+1} \sigma_i^x+\frac{1}{2}\sigma_1^x \right]\\
&=\left[(J^{1q}_1|_{\Delta,g\rightarrow N\Delta,Ng}+2g\mathcal{P}\sigma_1^z)\prod_{j=2}^{N+1} \sigma_j^z,a^\dagger a+\Delta \sum_{i=1}^{N} \sigma^z_i+g(a^\dagger+a)\sum_{i=1}^{N} \sigma_i^x+\frac{1}{2}\sigma_1^x \right]\\
&\quad +\left[(J^{1q}_1|_{\Delta,g\rightarrow N\Delta,Ng}+2g\mathcal{P}\sigma_1^z)\prod_{j=2}^{N+1} \sigma_j^z,\Delta \sigma^z_{N+1}+g(a^\dagger+a) \sigma_{N+1}^x \right]\\
&=\left[J^{1q}_1|_{\Delta,g\rightarrow N\Delta,Ng}\prod_{j=2}^{N} \sigma_j^z,a^\dagger a+\Delta \sum_{i=1}^{N} \sigma^z_i+g(a^\dagger+a)\sum_{i=1}^{N} \sigma_i^x+\frac{1}{2}\sigma_1^x \right]\sigma_{N+1}^z\\
&\quad +\left[2g\mathcal{P}\prod_{j=1}^{N+1} \sigma_j^z,\frac{1}{2}\sigma_1^x \right] +\left[J^{1q}_1|_{\Delta,g\rightarrow N\Delta,Ng}\prod_{j=2}^{N+1} \sigma_j^z,g(a^\dagger+a) \sigma_{N+1}^x \right].
\end{split}
\end{equation}
In the last line we have omitted vanishing terms. 
We can see that the first term cancels the first term in (\ref{dicketerm1finalform}) when summing them, 
so we only need to simplify the remaining two terms. 
With the commutation relations $\mathcal{P}a^\dagger=-a^\dagger \mathcal{P}$ and $\mathcal{P}a=-a\mathcal{P}$ in mind, 
we can compute the last two terms in (\ref{dicketerm2}), namely
\begin{equation}
    \begin{split}
    \left[ 2g\mathcal{P}\prod_{j=1}^{N+1} \sigma_j^z,\frac{1}{2}\sigma_1^x \right]=g\mathcal{P}\prod_{j=2}^{N+1} \sigma_j^z \left[ \sigma_1^z,\sigma_1^x \right]=2g\mathcal{P}\prod_{j=2}^{N}\sigma_j^z ( \mathrm{i} \, \sigma_1^y\sigma_{N+1}^z)
    \end{split}
\end{equation}
and
\begin{equation}
    \begin{split}
    &\left[J^{1q}_1|_{\Delta,g\rightarrow N\Delta,Ng}\prod_{j=2}^{N+1} \sigma_j^z,g(a^\dagger+a) \sigma_{N+1}^x \right]\\
    &=\prod_{j=2}^{N} \sigma_j^z\left[\mathcal{P}\left((a^\dagger+a)i\sigma_1^y+\frac{\Delta}{g}+(a^\dagger-a+2Ng)\sigma_1^x\right)\sigma_{N+1}^z,
   g(a^\dagger+a) \sigma_{N+1}^x \right]\\
   &=g\prod_{j=2}^{N}\sigma_j^z \left[\mathcal{P}(a^\dagger-a)\sigma_1^z\sigma_{N+1}^z,(a^\dagger+a) \sigma_{N+1}^x \right]=-2g\mathcal{P}\prod_{j=2}^{N}\sigma_j^z( \mathrm{i} \, \sigma_1^z\sigma_{N+1}^y).
    \end{split}
\end{equation}
Putting them together we have (\ref{dicketerm2}) rewritten as 
\begin{equation}\label{dicketerm2finalform}\begin{split}
& \left[J^{1q}_1 |_{\Delta,g\rightarrow (N+1)\Delta,(N+1)g}\prod_{j=2}^{N+1} \sigma_j^z,a^\dagger a+\Delta \sum_{i=1}^{N+1} \sigma^z_i+g(a^\dagger+a)\sum_{i=1}^{N+1} \sigma_i^x+\frac{1}{2}\sigma_1^x \right]\\
&= \left[J^{1q}_1|_{\Delta,g\rightarrow N\Delta,Ng}\prod_{j=2}^{N} \sigma_j^z,a^\dagger a+\Delta \sum_{i=1}^{N} \sigma^z_i+g(a^\dagger+a)\sum_{i=1}^{N} \sigma_i^x+\frac{1}{2}\sigma_1^x \right]\sigma_{N+1}^z\\
&\quad +2g\mathcal{P}\prod_{j=2}^{N}\sigma_j^z ( \mathrm{i} \, \sigma_1^y\sigma_{N+1}^z)-2g\mathcal{P}\prod_{j=2}^{N}\sigma_j^z(\mathrm{i} \, \sigma_1^z\sigma_{N+1}^y).
\end{split}
\end{equation}
Now with both terms in (\ref{dickeNp1qubit}) simplified, we can combine (\ref{dicketerm1finalform}) and (\ref{dicketerm2finalform}) 
to check that it indeed vanishes:
\begin{equation}\label{dickeNp1qubitvanish}\begin{split}
  & \left[J^{(N+1)q}, H^{(N+1)q} \right]\\
  & ~~~ =-\left[J^{1q}_1|_{\Delta,g\rightarrow N\Delta,Ng}\prod_{j=2}^N \sigma_j^z, \, a^\dagger a+\Delta \sum_{i=1}^N \sigma^z_i+g(a^\dagger+a)\sum_{i=1}^N \sigma_i^x +\frac{1}{2}\sigma_1^x \right]\sigma_{N+1}^z  \\
 &~~~~~  +2g\mathcal{P}\prod_{m\neq j}^{N}\sigma_m^z \mathrm{i} (\sigma_1^z\sigma_{N+1}^y-\sigma_1^y\sigma_{N+1}^z)\\
 &~~~~~ + \left[J^{1q}_1|_{\Delta,g\rightarrow N\Delta,Ng}\prod_{j=2}^{N} \sigma_j^z, \, a^\dagger a+\Delta \sum_{i=1}^{N} \sigma^z_i+g(a^\dagger+a)\sum_{i=1}^{N} \sigma_i^x+\frac{1}{2}\sigma_1^x \right]\sigma_{N+1}^z\\
&~~~~~ +2g\mathcal{P}\prod_{j=2}^{N}\sigma_j^z (\mathrm{i}\,\sigma_1^y\sigma_{N+1}^z)-2g\mathcal{P}\prod_{j=2}^{N}\sigma_j^z(\mathrm{i} \, \sigma_1^z\sigma_{N+1}^y)\\&=0.
\end{split}
\end{equation}

We have thus shown that the commutator between the $J$ operator and the Hamiltonian will vanish for the ($N$+1)-qubit system if it vanishes for 
the $N$-qubit system, and we know that it vanishes for the 1-qubit system. 
Thus by induction, we conclude that the $J$ operator (\ref{JoperatorDef}) commutes with the Hamiltonian (\ref{BDMHamiltonian}) 
for arbitrary number of qubits and hence generates a symmetry for the $N$-qubit biased Dicke model.

\section{Proof of Theorem 2}

We will also prove this theorem by induction. 
Again, we set $\epsilon_1=\frac{1}{2}$ for simplicity. 
Assuming the $N$-qubit case is true, then the ($N$+1)-qubit case is also true provided 
\begin{equation}\label{JSquareDif}
   (J^{(N+1)q})^2=(J^{Nq})^2+4(\Delta \sigma_{N+1}^z+g(a^\dagger+a)\sigma_{N+1}^x)+4(2N+1)g^2-16g^2\sum_{j=2}^N B_{j,N+1}.
\end{equation}
Our goal here is to prove this equation.

From (\ref{JoperatorDef}) we can rewrite $J^{(N+1)q}$ in terms of $J^{Nq}$,
\begin{equation}
    J^{(N+1)q}=J^{Nq}\sigma_{N+1}^z+2g\mathcal{P}\prod_{n=1}^{N+1}\sigma_n^z+4g\mathcal{P}B_{1,N+1}\prod_{j=2}^N \sigma_j^z.
\end{equation}
Calculating its square gives
\begin{equation}
    \begin{split}\label{JSquareStart}
    (J^{(N+1)q})^2&=(J^{Nq})^2+2gJ^{Nq}\mathcal{P}\left(\prod_{n=1}^{N}\sigma_n^z+2\sigma_{N+1}^z B_{1,N+1}\prod_{j=2}^N \sigma_j^z\right)\\
    &\phantom{=}+2g\mathcal{P}\left(\prod_{n=1}^{N}\sigma_n^z+2 B_{1,N+1}\prod_{j=2}^{N+1} \sigma_j^z\right)J^{Nq}+4g^2\\
    &\phantom{=}+8g^2\sigma_1^z\sigma_{N+1}^zB_{1,N+1}+8g^2B_{1,N+1}\sigma_1^z\sigma_{N+1}^z+16g^2(B_{1,N+1})^2.
    \end{split}
\end{equation}
By using the properties (\ref{prop1}) and (\ref{prop2}) the last line of (\ref{JSquareStart}) is simply $16g^2B_{1,N+1}$. 
Then we only need to expand the two brackets, \textit{i.e.}, the second and the third terms. 
We begin with the second term. 
Using the definition of the $J$ operator (\ref{JoperatorDef}), we first expand $J^{Nq}$, 
\begin{equation}
\begin{split}
&J^{Nq} \mathcal{P}\left( \prod_{n=1}^{N}\sigma_n^z+2\sigma_{N+1}^z B_{1,N+1}\prod_{j=2}^N \sigma_j^z \right) \\
&=\mathcal{P}\left((a^\dagger+a) \mathrm{i} \sigma_1^y+\frac{\Delta}{g}+(a^\dagger-a+2Ng)\sigma_1^z \right)\prod_{j=2}^N\sigma_j^z \mathcal{P}\left(2\sigma_{N+1}^z B_{1,N+1}\prod_{j=2}^N \sigma_j^z\right. \\
&\qquad \left.+\prod_{n=1}^{N}\sigma_n^z\right) +4g\mathcal{P}\sum_{i=2}^N B_{1i}\prod_{j\neq i}^N \sigma_j^z\mathcal{P}\left( \prod_{n=1}^{N}\sigma_n^z+2\sigma_{N+1}^z B_{1,N+1}\prod_{j=2}^N \sigma_j^z \right)\\
&=\left(-(a^\dagger+a) \mathrm{i} \sigma_1^y+\frac{\Delta}{g}+(-a^\dagger+a+2Ng)\sigma_1^z \right)\left( \sigma_1^z+2\sigma_{N+1}^z B_{1,N+1} \right)\\
&\quad +4g\sum_{i=2}^N B_{1i} \sigma_1^z\sigma_i^z+8g\sum_{i=2}^N B_{1i}\sigma_i^z\sigma_{N+1}^z B_{1,N+1}.
\end{split}
\end{equation}
We see that in the second line above, $\prod_{j=2}^{N}\sigma_j^z$ commutes with everything, 
so we can move it around to cancel a similar factor as it squares to the identity. 
Similarly, using commutation relations between $\mathcal{P}$ and $a^\dagger,a$ 
we can bring the $\mathcal{P}$ operators together as they also square to the identity. 
These steps result in the last line of the expression above.

We can do exactly the same thing for the third term in (\ref{JSquareStart}), resulting in 
\begin{equation}
\begin{split}
\mathcal{P}&\left(\prod_{n=1}^{N}\sigma_n^z+2 B_{1,N+1}\prod_{j=2}^{N+1} \sigma_j^z\right)J^{Nq} \\
&=\left( \sigma_1^z+2 B_{1,N+1}\sigma_{N+1}^z \right)\left((a^\dagger+a) \mathrm{i} \sigma_1^y+\frac{\Delta}{g}+(a^\dagger-a+2Ng)\sigma_1^z \right)\\
&\quad +4g\sum_{i=2}^N \sigma_1^z\sigma_i^z B_{1i} +8gB_{1,N+1}\sigma_{N+1}^z\sum_{i=2}^N \sigma_i^z B_{1i} .
\end{split}
\end{equation}
Note here the relative positions of $B$ and $\sigma^z$ are different to the second term.

With these two simplified expressions, it is much easier to calculate their sum. 
Starting with the brackets containing $a^\dagger$ and $a$, without much surprise, 
most of them cancel, leaving the terms
\begin{equation}
    2(a^\dagger+a)\sigma_{N+1}^x+2\frac{\Delta}{g}\sigma_{N+1}^z+4Ng-8NgB_{1,N+1} .
\end{equation}
As for the remaining terms, namely
\begin{equation} \label{terms}
   4g\sum_{i=2}^N B_{1i} \sigma_1^z\sigma_i^z+4g\sum_{i=2}^N \sigma_1^z\sigma_i^z B_{1i}+8g\sum_{i=2}^N B_{1i}\sigma_i^z\sigma_{N+1}^z B_{1,N+1} +8gB_{1,N+1}\sigma_{N+1}^z\sum_{i=2}^N \sigma_i^z B_{1i},
\end{equation}
the first two terms reduce to $-8g\sum_{i=2}^N B_{1i}$. 
However, the last two terms are non-trivial.
To simplify them, we need the $B$-matrix property (\ref{prop3}).
This property simplifies the last two terms in (\ref{terms}) to 
\begin{equation}
    8g\sum_{i=2}^N(B_{1i}+B_{1,N+1}-B_{i,N+1}).
\end{equation}

We have thus simplified all terms in (\ref{JSquareStart}), which now reads  
\begin{equation}
    \begin{split}
    (J^{(N+1)q})^2&=(J^{Nq})^2+16g^2B_{1,N+1}+4g^2+2g\bigg( 2(a^\dagger+a)\sigma_{N+1}^x+2\frac{\Delta}{g}\sigma_{N+1}^z\\ &\quad+4Ng-8NgB_{1,N+1}-8g\sum_{i=2}^N B_{1i}+8g\sum_{i=2}^N(B_{1i}+B_{1,N+1}-B_{i,N+1})\bigg)\\
    &=(J^{Nq})^2+4(\Delta \sigma_{N+1}^z+g(a^\dagger+a)\sigma_{N+1}^x)+4(2N+1)g^2-16g^2\sum_{i=2}^N B_{i,N+1}.
    \end{split}
\end{equation}
By comparing this resulting expression with (\ref{JSquareDif}), we see that they are exactly the same. 
It follows that the relation (\ref{quadraticRelationJandH}) is true for the $N$-qubit biased Dicke model. 
It further follows that the hidden symmetry generated by $J$-operator (\ref{JoperatorDef}) is a $\mathbb{Z}_2$ symmetry.

\end{document}